\documentclass[preprint,showpacs,preprintnumbers,amsmath,amssymb,floats]{revtex4}
\usepackage[dvips]{graphicx}
\usepackage[english]{babel}
\usepackage{amsmath}
\usepackage{amssymb}
\usepackage{color}

\newcommand{\eijk}{e_{ijk}}

\newcommand{\be}{\begin{eqnarray}}
\newcommand{\ee}{\end{eqnarray}}

\begin{document}

\title{Gyrotropic Birefringence in the Under-doped Cuprates}
\author{C. M. Varma}
\affiliation{Physics Department, University of California, Riverside, Ca. 92521}
\date{\today}

\begin{abstract}
The optical effects due to the loop-current order parameter in under-doped cuprates are studied 
in order to understand the recent observation of unusual birefringence in electromagnetic propagation in under-doped cuprates.
It is shown why birefringence occurs even in multiple domains of order with size  of domains much smaller than the wave-length and in twinned samples. Not only is there a rotation of polarization of incident light but also a rotation of the principal optical axis from the crystalline axes. Both are calculated in relative agreement with experiments in terms of the same parameters. The magnitude of the effect is orders of magnitude larger than the unusual Kerr effect observed in under-doped cuprates earlier.  The new observations, including their comparison with the Kerr effect, test the symmetry of the proposed order decisively and confirm the conclusions from polarized neutron scattering.
\end{abstract}
\maketitle

{\it Introduction}: 
The heart of the solution of the cuprate problem is in ascertaining the physics of the strange metal phase \cite{mfl} and the so-called pseudo-gap phase \cite{norman-pines-kallin}, \cite{timusk-rev}. One theoretical approach  \cite{loop-order}, \cite{cmv-2006}, \cite{Aji}  suggests that the strange-metal region of the cuprates  as well their high temperature d-wave superconductivity is due to the scattering of fermions from the quantum-critical fluctuations of an unusual time-reversal breaking phase, which has a finite magneto-electric tensor and which occupies the so-called pseudo-gap region of the phase diagram. Various experiments \cite{He-arpes-kerr}, \cite{leridon}, \cite{ultrasound}  have provided evidence for a broken symmetry at the onset of the pseudo-gap region. Polarized neutron scattering in four families of cuprates \cite{loop-order-expt}
 and dichroic ARPES \cite{kaminski} in one have provided direct evidence of the predicted symmetry. Controversy, some of it based on scientific grounds \cite{nmr} \cite{muons} \cite{cmv-pseudogap}, however continues.
It is therefore of great interest that experiments \cite{armitage} with quite a different technique are now available to address the issue of the phase transition to the pseudo-gap phase and the specific symmetry of the magneto-electric tensor proposed for it.

These experiments are optical experiments \cite{armitage} which have observed birefringence with several unusual features in under-doped cuprates below the temperatures $T^*(x)$ at which other signatures of the pseudo-gap are observed. The fact that they begin to be observed below $T^*(x)$ and their magnitude grows below it indicates of-course that $T^*(x)$  
marks a new symmetry. But the experiments are done in twinned films so that birefringence would normally not be observed. Even in single-crystals, no birefringence can be usually observed if there is a symmetry breaking but with domains of different equivalent order of size much smaller than the optical wavelength. The other remarkable features of the observations are that the principal axes of the birefringence are not the crystalline axes but rotated by an angle $\theta_{P}$ with respect to them. Another feature is that the angle of rotation of the polarization $\theta_R$ and $\theta_P$  defined with respect to the propagation direction remains the same on shining light on the two opposite faces of the sample with respect to the a-b planes.  In materials with time-reversal breaking, some unusual optical effects have been predicted \cite{Brown} and some observed but none of the kind discovered in cuprates. I show here that the observations follow from the symmetry of the magneto-electric tensor proposed to occur in the cuprates. I also suggest further experiments to verify some untested aspects of the results obtained here. I will also compare the difference in the occurrence and magnitude of the unusual Kerr effect \cite{kerr-ybco}, \cite{kerr-more} observed earlier in under-doped cuprates compared to the recent birefringence measurements. There is much to be learnt from this comparison.

 The general optical effects in materials with magneto-electric symmetry and with application to the symmetry of Cr$_2$O$_3$ and MnTiO$_3$  were proposed by Brown et al. \cite{Brown}; the corresponding Maxwell equations were written down by Hornreich and Shtrikman \cite{HS}. I write them here for the propagation of light in symmetry appropriate to that proposed for the cuprates and provide a physical explanation for the unusual features.
 
{\it Analysis}: Let us start with the Maxwell Equations:
\be
\label{ME}
c \eijk E_j q_k= -\omega B_i, ~~c\eijk H_j q_k = \omega D_i.
\ee
Where $e_{ijk}$ is the totally anti-symmetric unit matrix. Following Agranovich and Ginzburg \cite{AG}, one can combine all induced effects in a generalized polarizability $P'_i$ rather than introducing them separately in an induced magnetization by defining 
\be 
P_i' = P_i + (c/\omega) \eijk M_j q_k, ~~ D'_i = E_i + 4 \pi P'_i
\ee
so that the Maxwell Equations can be written as 
\be
\label{ME'}
c \eijk E_j q_k= -\omega B_i, \\
c\eijk H_j q_k = \omega D'_i.
\ee
The material properties are defined through the relations
\be
\label{const rel}
D_i = \epsilon_{ij} E_j + \chi^{EM}_{ij}H_j,\\
B_i = \chi^{ME}_{ij} E_j + \mu_{ij} H_j.
\ee
$ \chi^{EM}_{ij}$ and $ \chi^{ME}_{ij}$ are the elements of the magneto-electric tensor. It is assumed that there is no natural optical activity in the material, i.e. the material is not dielectrically chiral. The effective polarization $D'_i$ may then be written with $(i,j,k)$ taken as the principal axes such that  the ordinary dielectric matrix $\epsilon_{ij}$ and the permittivity matrix $\mu_{ij}$ are diagonal, with values $\epsilon_{i}$ and  $\mu_i$ respectively. Then
\be
\label{effD}
D'_i = \Big(\bar{\epsilon} _{ij}  + \bar{\gamma}_{ijk} q_k -(c/\omega)^2 (1- \mu^{-1}_i)q_k^2\delta_{ij}(1-\delta_{ik}) \Big)E_j
\ee
where the renormalized susceptibility and gyrotropic tensors are
\be
\label{reno}
\bar{\epsilon} _{ij} = \epsilon_i \delta_{ij} - \chi^{EM}_{ik}\mu^{-1}_k\chi^{ME}_{kj},\\
\bar{\gamma}_{ijk} = i(c/\omega)(e_{ijl}\chi^{ME}_{lk} +  e_{jkl} \chi^{EM}_{il} )\mu^{-1}_{l}.
\ee

 \begin{figure}
\centerline{\includegraphics[width=0.9\textwidth]{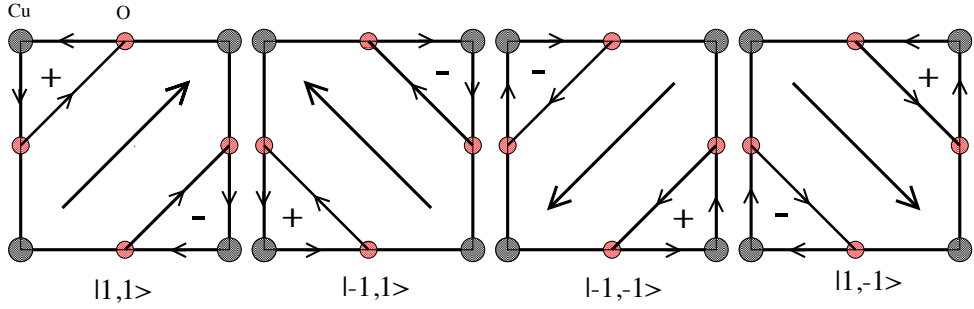}}
\caption{The four Possible  domains of the loop ordered state are shown. The current loops in each and the resulting direction of magnetic moment is shown. The direction of the anapole vector is also shown and labelled by $|\pm1,\pm1>$. Their magneto-electric tensor is specified in the text.}
\label{domains}
\end{figure}

{\it Effects in specific Symmetry of the Loop Ordered State}: We now consider the symmetry of the proposed state in underdoped cuprates where an order parameter exists so that 
$\chi^{ME}_{ij}, \chi^{EM}_{ij}$ are finite. This order parameter arises due to orbital currents which have two current loops in each unit-cell as depicted in Fig. (\ref{domains}). The order parameter is characterized by the anapole vector
\be
{\bf \Omega} = \int_{cell} d^2r ({\bf M}({\bf r}) \times {\bf r}).
\ee
In a tetragonal crystal the symmetry class is $(\underbar{m}mm)$ \cite{birss}. I will assume a tetragonal symmetry even in orthorhombic crystals for calculation of the extra effects due to the loop current order because the correction due to the orthorhombic symmetry are small. (We must however keep track of the ordinary birefringence of the orthorhombic symmetry.)  There are four domains, two with  anapole vector oriented in the $\pm \hat{\bf x}' = \pm (1/\sqrt{2})(\hat{\bf x} +\hat{\bf y})$ directions, and two with anapole vectors oriented in the $\pm \hat{\bf y}' = \pm (1/\sqrt{2})(\hat{\bf x} -\hat{\bf y})$ directions. The former have only non-zero real elements $\chi^{EM}_{x',z} = \chi^{ME}_{z,x'}$ \cite{birss}, with equal values for the two domains. Similarly the latter have only non-zero real elements $\chi^{EM}_{y',z} = \chi^{ME}_{z,y'}$. 
Also $\chi^{EM}_{y',z}  = \chi^{EM}_{x',z}$. These are evident already from the cartoons in Fig. (\ref{domains}).

We will consider here only propagation in the z-direction: ${\bf k} = k_z$. In this case the relevant non-zero off-diagonal elements of the renormalized permittivity are  
\be
\bar{\epsilon}_{xy} =  \mu^{-1}_z \chi^{EM}_{xz} \chi^{ME}_{zy}=
\bar{\epsilon}_{yx}. 
\ee
There is also an equal corrections to the diagonal parts $\epsilon_{xx}$ and $\epsilon_{yy}$ which may be ignored due to the much larger ordinary parts.
The only components of $\gamma$ that are relevant for $\hat{z}$-axis propagation are $\gamma_{xyz}$ and $\gamma_{yxz}$. From Eq. (\ref{reno}), it follows that they are 0.  

The occurrence of an off-diagonal component in the dielectric permittivity $\bar{\epsilon}_{xy}$, from which all effects derived here follow, starting from a magneto-electric tensor whose components are $\chi^{ME}_{xz}$ etc., may be understood as follows. The Free-energy contains terms of the form $(\chi^{ME}_{xz} + \chi^{EM}_{zx
})H_zE_x$ (plus similar terms with $x \to y$) as well as a term $H^2/\mu_z$. On propagation of light along the $z$-axis, i.e. an electric field in the 
$x$-direction generates a magnetic field in the $z$-direction which may be eliminated and an effective Free-energy term $\bar{\epsilon}_{xy} E_x E_y$ obtained with $ \bar{\epsilon}_{xy}  = \chi^{ME}_{xz}\mu_z^{-1}\chi^{EM}_{zy}$. This quadratic dependence of $\bar{\epsilon}_{xy}$ on the order parameter plays an important role in what follows.

Inserting Eq. (\ref{reno}) in (\ref{effD}) and considering transverse propagation alone and using the Maxwell equations (\ref{ME'}),
\be
\label{propcond1}
D'_x =  (c/\omega)^2 k_z^2 E_x=  (\epsilon_{xx} +(c/\omega)^2 k_z^2(1-\mu^{-1}_x))E_x + (\bar{\epsilon}_{xy})E_y\\
\label{propcond2}
D'_y = (c/\omega)^2 k_z^2 E_y= (\epsilon_{yy} +(c/\omega)^2 k_z^2(1-\mu^{-1}_y))E_y + (\bar{\epsilon}_{yx} )E_x
\ee
The usual refractive indices are given by $n^2_{0x} = \epsilon_x\mu_x, n^2_{0y} = \epsilon_y\mu_y$. Rewrite Eqs. (\ref{propcond1}, \ref{propcond2}) in terms of the allowed refractive indices $n \equiv (ck_z/\omega)$ and $n^2_{xy}\equiv \mu_x \bar{\epsilon}_{xy}, n^2_{yx}\equiv \mu_y \bar{\epsilon}_{xy}$.
\be
\label{ans1}
(n^2_{0x}- n^2) E_x + n^2_{xy} E_y=0, \\
\label{ans2}
n^2_{yx} E_x + (n_{0y}^2-n^2) E_y=0.
\ee
Eqs.(\ref{ans1}, \ref{ans2}) give that the solution for the n's is,
\be
\label{n}
(n_1^2, n_2^2) = 1/2(n_{0x}^2 + n_{0y}^2) \pm  1/2 \sqrt{ (n_{0x}^2 - n_{0y}^2)^2 + 4\Lambda^4 }.
 \ee
where $\Lambda^2 = n_{xy}n_{yx}$. 
 The eigenvectors of (\ref{ans1},\ref{ans2}) have the property that 
  \be
 \label{ex/ey}
 \big(E_x/E_y\big)_1 & = & \Lambda^2/(n^2_1-n^2_{ox}), \\ \nonumber
 \big(E_y/E_x\big)_2 & = & \Lambda^2/(n^2_2 -n^2_{oy}).
 \ee
These specify the two principal optical axes, i.e when the incident polarizations has these ratios, there is no dichroism. For all other angles there is.

Now we may explore the consequences of Eqs. (\ref{n}, \ref{ex/ey}).  The first important result is that since $\Lambda^4$ is proportional to $\bar{\epsilon}_{xy}^2$, the results are independent of domains of order in general. Some specific results are different for single-crystals and poly-crystalline materials, as described below. The former must further be divided into single-domain order (which may be hard to obtain) or order with many domains in the field of vision. 

{\it Single Crystals} \\
It follows from (\ref{ex/ey}) that in a tetragonal crystal, i.e. $n_{ox} = n_{0y}= n_0$,  the principal axes are at $\pi/4, 3\pi/4$ to the crystalline axes.
However, the velocity of propagation when the polarization is at any other angle is increased/decreased for the projection to these principal axes,
\be
\label{veldiff}
(c/n_{1}, c/n_2) \approx \pm c/n_0\big(1 \mp |\Lambda^2|/n_0^2\big)
\ee
Changing the propagation from $+\hat{z}$-direction to $-\hat{z}$-direction is equivalent to redefining the axes: $\hat{x} \to \hat{x}; \hat{y} \to -\hat{y}$, or the other way around.  This is equivalent to changing $n_1 \to n_2$. So,  as in ordinary birefringence, their is no rotation in propagation in a given direction and then back along the same path.  For order with multiple domains in single crystals, which is to be expected, the effects remain the same because $|\Lambda^2|$ is the same in all four domains. The switching of the principal axes from the crystalline axes for $T \gtrsim T^*(x)$ to half-way between them for $T \lesssim T^*(x)$ is a strong prediction of the considerations here. As usual effects due to impurities, fluctuations etc., will in general round out the transition.

For an orthorhombic single crystal, Eq. (\ref{ex/ey}) gives that the rotation-angle of the principal axes $\theta_P \approx \arctan(|\Lambda^2|/(n_{0x}^2 - n_{0y}^2))$. The value of the rotation angle, $\theta_R$  given by  for arbitrary polarization  may be directly determined by projecting the initial polarization to the principal axes and noting the change in polarization in propagation due to the difference in velocities of the two components, (\ref{veldiff}). So there is both a rotation of the principal axes, $\theta_{P} \ne 0$ and rotation of polarization $\theta_R \ne 0$. These effects remain the same due to multiple domains of order in a single orthorhombic crystal because $|\Lambda^2|$ remains the same. Both effects are in general temperature dependent because $\Lambda^2$
increases for $T \leqslant T^*(x)$

{\it Twinned Samples}

The available experiments  are performed in twinned samples of the orthorhombic compound YBa$_2$Cu$_3$O$_{6+x}$ with twin size much smaller than the wavelength of light. In this case, a suitable averaging procedure must be found. Consider Eq. (\ref{n}). Obviously, in the twinned samples, the first term $1/2(n_{0x}^2 + n_{0y}^2)$ may be averaged to its mean value $\bar{n}^2$. But the second term does not average to zero for $\Lambda^2 \ne 0$.
As we will discuss below, the measured $(\delta n^2) \equiv |n_{0x}^2 - n_{0y}^2|$ is expected to be much larger than $\Lambda^2$. So, we may approximate,
\be
\label{nbar}
(\bar{n_1^2}, \bar{n_2^2}) &\approx & \bar{n^2} \pm  2 \Lambda^4/\delta n^2,\\
E_x/E_y &=& E_y/E_x = \pm \frac{2|\Lambda^2|}{\overline{\sqrt{((\delta n^2)^2) + 4 \Lambda^4}}} \approx  \pm \frac{2\Lambda^2}{(\delta n^2)}.
\ee
This gives that the principal axes are at
\be
\label{p.a.}
{\bf\hat{d_1}} \approx\big({\bf\hat{x}} + \frac{2|\Lambda^2|}{(\delta n^2)}{\bf\hat{y}}\big),
 \ee
 to $O(\Lambda^4 /(\delta n^2)^2$, and the vector ${\bf\hat{d_2}}$ orthogonal to it and the directions opposite to ${\bf\hat{d_1}},{\bf\hat{d_2}}$.
 
 Consider the magnitude of the dichroism. It is given in terms of the ratio of the difference of the velocities to the average velocity,\\
 \be
 \label{velocities}
 \delta v /v \equiv c(1/n_1-1/n_2)/(c/\bar{n}) \approx 1/2 \big(\overline{\sqrt{(\delta (n^2)^2) + 4 \Lambda^4}}\big)/\bar{n}^2 \approx \frac{\Lambda^4}{(\delta n^2)}\frac{1}{\bar{n^2}}.
 \ee
The angle of rotation (which varies in a four-fold way with respect to the angles $\theta_P$) has as its maximum in units of angular rotation per travel over a wave-length,
\be
\theta_{Rm} \approx 4\pi \frac{\Lambda^4}{(\delta n^2)}\frac{1}{\bar{n^2}} ~Radians.
\ee
As temperature decreases, the magnitude of birefringence should increases as the fourth power of the loop-current order parameter. In experiments, the dichroism is expressed in terms of a complex angle $\theta_R = \theta'_{R} + i \theta"_{R}$. The rotation angle is complex because typically in a metal the real and imaginary parts of the index of refraction are equal because $\epsilon" \gg \epsilon'$. In experiments $\theta'_{R} \approx \theta"_{R}$. This is of-course obtained from Eq. (\ref{velocities}) for such a normal complex refractive index $\bar{n}$.
 
 {\it Comparison with Experiments}
 
Let us first consider the unusual qualitative features of the experimental results \cite{armitage}.
 (1) A birefringence is observed in a {\it polycrystal} below $T \approx T^*(x)$ and increases in magnitude as temperature is decreases, (2) the principal axes are rotated with respect to the crystalline axes, and (3) the effect has the same sign for shining light normally on opposite basal plane faces of the sample. All these three remarkable effects follow from Eqs. (\ref{p.a.}, \ref{velocities}). 
They arise because $|\Lambda^2|$ is independent of switching of the x and the y-axis and of the orientation of the domains of ${\bf \Omega}$ as shown above. Note that the observed effects test the particular symmetry predicted of the magneto-electric phase. The results are quite different from what is predicted and observed for example in the classic antiferromagnetic-magneto-electric insulator Cr$_2$O$_3$ primarily because its magneto-electric tensor is diagonal while it is off-diagonal, as specified above, for the under-doped cuprates. Polarization with an external electric and/or Magnetic field to align the domains of order in a single crystal of Cr$_2$O$_3$ are therefore required to observe $\theta_P$ and $\theta_R$. 

From Eq.(\ref{velocities}), it follows that, since $\Lambda^2$ is proportional to the square of the order parameter, the rotation angle $\theta_R$ should increase as the fourth power of the order parameter. I note that in the experiment, the increase of $\theta_R$ below an approximately determined $T^*(x)$ may not be fitted as a linear increase but may be fitted as $(T^*(x)-T)^2$, consistent with the above result for a mean-field value of the order parameter exponent. 
 
The experiments \cite{armitage} give both the real and imaginary parts of $\theta_R$ as a function of temperature and report that at low temperatures $\theta_P \approx 10~ degrees$. They do not report the temperature dependence of $\theta_P$.  A temperature dependence is predicted by Eq. (\ref{p.a.}). Direct quantitative comparison with the experimental results is  hampered by the lack of information on $(\delta n^2)$ at the measurement frequency and lack of an absolute determination of $\Lambda^2$. The dimensionless order parameter, the ratio of the imaginary component of the transfer integral induced by time-reversal breaking to the normal real part around a loop, is estimated \cite{cmv-2006}, \cite{weber-2013} to be $O(10^{-1})$. But it is not possible to reliably calculate the magnitude of the $\epsilon_{xy}$ from this, (nor am I aware of instances where this kind of thing has been calculated reliably in even much simpler situations where usually only the symmetry of the optical anisotropies are tested in experiments). 

We may however relate the parameters required to get the measured value of $\theta_P$ at low temperatures to the measured value of $\theta_R$. We can use Eq. (\ref{p.a.}) to find $\Lambda^2/(\delta n)^2 \approx 1/12$ for the measured $\theta_P$ of about 10 degrees at the lowest temperature.  The difference in conductivity \cite{Cooper}, \cite {Wang} of single crystal YBa$_2$Cu$_3$O$_{6.6}$ at frequencies down to about $10^{13}$ Hz, (an order of magnitude larger than the birefringence measurements), along the crystalline axis  may be read off to conclude that  due to the conductivity of the chains, $(\delta n^2)/n^2$ is of $O(1)$. If we use this estimate, we get that from (\ref{velocities}) that the  rotation angle should be about $80~ mRadians$, while the experiments \cite{armitage} give a maximum of about $60 ~mRadians$. 

Finally, I contrast the measurements of the unusual (Kapitulnik) Kerr effect \cite{kerr-more} \cite{kerr-ybco} in under-doped YBa$_2$Cu$_3$O$_{6+x}$ with the 
gyrotropic birefringence. Several explanations of the Kerr effect have been proposed \cite{kerr-theories}. In single-crystals of this compound, Kerr effect occurs starting at a temperature $T_{KK}(x) < T^*(x)$ (but extrapolating to $T \to 0$ to the quantum-critical point at $x_c$.) The birefringence appears to occur starting at  $T^*(x)$ measured by transport and by polarized neutrons. It has been shown earlier \cite{kerr-ahv} that the magneto-electric state does not have a Kerr effect, but that given its presence, lattice distortions of certain symmetry necessarily induce an additional loop order which has an anomalous Hall effect and therefore a Kerr effect. Such lattice distortions do appear to occur in YBa$_2$Cu$_3$O$_{6+x}$ at temperatures consistent with $T_{KK}(x)$. The magnitude of the Kerr effect then depends on an additional small parameter - the square of the relevant lattice distortion and is therefore expected to be much smaller than the birefringence. Because the experiments are done at quite different frequencies, direct comparison is not possible. But it should be noted that the angle in similar units is about four orders of magnitude smaller in the Kerr effect. It should also be noted that one of the peculiar features of the Kerr effect is that it is observed at all - since domains of differently oriented order of size much smaller than the wave-length are expected to be present in such a sample. Moreover, the direction of rotation is independent of heating the sample across $T_{KK}(x)$ and then measuring on cooling below. These unusual features have been uniquely explained earlier \cite{kerr-ahv}
as also from some of the same features of the loop current order as used above for the birefringence. Note also that we found that the Kerr effect should increase below $T_{KK}(x)$ proportionally to the square of the induced order parameter, in contrast to the fourth power of the order parameter for the birefringence obtained above. The experiments are consistent with this difference.

The  results obtained here can be further tested by doing birefringence experiments in single-crystals, both in tetragonal and orthorhombic symmetries and at the frequencies where Kerr effect results are available. In particular a different magnitude of the effects and their temperature dependence is predicted than for poly-crystals.

I am grateful to N.P. Armitage and Y. Lubashevski for a detailed discussion of the experimental results and to Vivek Aji and J. Orenstein for very useful 
remarks.

\end{document}